\documentstyle[epsfig,eqsecnum,twocolumn,aps,prd]{revtex}
\begin{document}
\draft
\title{How $\pi^0 \rightarrow \gamma \gamma$ 
changes with temperature}
\author{Robert D. Pisarski, T.L. Trueman and Michel H. G. Tytgat}
\address{Department of Physics, Brookhaven National Laboratory, 
Upton, New York 11973-5000, USA} 
\date{\today} 
\maketitle
\begin{abstract}
At zero temperature, in the chiral limit
the amplitude for $\pi^0$ to decay into
two photons is directly related to the coefficient of the axial anomaly.
At any nonzero temperature,
this direct relationship is lost:
while the coefficient of the axial anomaly is independent of
temperature, in a thermal bath
the anomalous Ward identities do
not uniquely constrain 
the amplitude for $\pi^0 \rightarrow \gamma\gamma$.
Explicit calculation shows that 
to lowest order about zero temperature, this amplitude decreases.
\end{abstract}
\pacs{12.39.Fe  11.10.Wx  11.30.Qc  11.30.Rd}
\section{Introduction}
\label{sec:intro}

In field theory, currents which are conserved classically
may not be quantum mechanically~\cite{adler}.  
For example, in massless
$QED$ the conservation of the axial current is violated by
the axial anomaly.  
The Adler-Bardeen theorem states that with the proper regularization scheme,
the coefficient of the anomaly, as computed at one loop order,
is exact to all orders in perturbation 
theory.  Moreover, the
axial anomaly does not change if the fermions propagate in
either a thermal bath or a Fermi sea~\cite{anom2,mueller,anom4}.  

In vacuum, one of the most 
striking manifestations of the axial anomaly
is the decay of a neutral pion into two photons, as the amplitude
is directly proportional to the coefficient of the 
axial anomaly in QED~\cite{adler,shore}.  
A natural supposition is then that because 
the axial anomaly does not change with
temperature, neither does the amplitude for 
$\pi^0 \rightarrow \gamma \gamma$~\cite{estrada1,estrada2}.  

In this paper we show that the story is  more involved.
We compute with a gauged nonlinear
sigma model~\cite{chiral} 
which properly incorporates all anomalies
by inclusion of the Wess-Zumino-Witten (WZW) 
term~\cite{wess-zumino,witten,kayma,spence,creutztyt}.
The effective lagrangian for $\pi^0 \rightarrow \gamma \gamma$ is
\begin{equation}
{\cal L}_{\pi\gamma\gamma} = \left ( {e^2 N_c \over 48\pi^2}\right )  
\; {1\over f_\pi} \;
\pi^0 F_{\alpha \beta} \widetilde{F}^{\alpha \beta} \; ,
\label{eq:1.1}
\end{equation}
where $f_\pi \sim 93 MeV$ is the pion decay constant, 
$N_c = 3$ is the number of colors, {\it etc.}

In sec.~\ref{sec:vac} we start by computing the effects
of pion loops on the amplitude of (\ref{eq:1.1}),
using the WZW action to one loop order in 
vacuum~\cite{wzwloop}.
The form of the WZW action is constrained by topology~\cite{witten},
so after the dust of calculation settles, in vacuum the result
is trivial: the only effect of the pion loops is to change
a bare pion decay constant into a renormalized $f_\pi$.

We then extend the calculations to
soft, cool pions at low temperature~\cite{pistyt2}.  
In this paper we work exclusively with two
flavors in the chiral limit.
The restriction to two flavors is done for
ease of calculation, and is otherwise inessential.
The chiral limit, $m_\pi = 0$,
is assumed because then the pion decay amplitude 
is directly related to the axial anomaly; for
calculations at $m_\pi \neq 0$ at nonzero temperature,
see~\cite{estrada2,piot}.
We believe that our results are relevant for $m_\pi \neq 0$
(as in vacuum), but more detailed analysis is
required to establish this.
 
At nonzero temperature, calculations in 
a background field formalism~\cite{pistyt2}
show that to $\sim T^2/f_\pi^2$,
the zero temperature pion decay constant is
replaced by a temperature dependent form~\cite{fpit,pistyt1},
\begin{equation}
f_\pi(T) = \left(1 - {1 \over 12} 
{T^2\over f_\pi^2}\right) f_\pi \; ,
\label{eq:1.2}
\end{equation}
Thus a second guess for the change of 
${\cal L}_{\pi\gamma\gamma}$ with temperature
would be that the zero temperature  $f_\pi$ is replaced
by $f_\pi(T)$.  Since $f_\pi(T)$ decreases to $\sim T^2/f_\pi^2$,
if true the amplitude, $\sim 1/f_\pi(T)$,
would increase to this order.

In sec.~\ref{sec:lowt} we evaluate precisely the same
diagrams as at zero temperature to $\sim T^2/f_\pi^2$.
The result, (\ref{eq:3.9}), is the sum of two terms:
one has exactly the form of (\ref{eq:1.1}), 
with $f_\pi$ replaced by $f_\pi(T)$, but
there is also a second term, special to nonzero temperature.
This type of term was derived recently in nonlinear
sigma models in the absence of gauge fields~\cite{pistyt2}: it is
nonlocal, analogous to the hard thermal loops of hot
gauge theories~\cite{htl}.  
For $\pi^0 \rightarrow \gamma \gamma$,
to order $\sim T^2/f_\pi^2$ we find that the sum of these
two terms is such that instead of
increasing, like $\sim 1/f_\pi(T)$, the amplitude
decreases, like $\sim f_\pi(T)$, (\ref{eq:3.8a}).

In sec.~\ref{sec:ward} we give a general analysis of the
relationship between the chiral Ward identities and
the amplitude for $\pi^0 \rightarrow \gamma \gamma$.
As is standard \cite{shore}, we use the anomalous Ward identity 
to relate a three point function of currents to the amplitude for
$\pi^0 \rightarrow \gamma \gamma$.
At zero temperature, this relationship is precise because
of the Sutherland-Veltman theorem 
\cite{suther-veltman} (in a slight abuse of
terminology).  The proof of the Sutherland-Veltman theorem
depends crucially upon Lorentz invariance.
A heat bath, however, provides a preferred rest frame;
extending an analysis of Itoyama and Mueller~\cite{mueller},
we show that consequently, the Sutherland-Veltman theorem
does not apply at any nonzero temperature.
This is why $\pi^0 \rightarrow \gamma \gamma$
changes with temperature, even though the anomaly doesn't:
besides the contributions from $\pi^0 \rightarrow \gamma \gamma$,
because there is no Sutherland-Veltman theorem 
at nonzero temperature, there are other terms which enter
to ensure that the Adler-Bardeen theorem is satisfied.

In sec.~\ref{sec:adlerbardeen} we demonstrate
these general arguments 
by computing the correlator between one axial
and two vector currents in the nonlinear
sigma model to one loop order, $\sim T^2/f_\pi^2$.
At nonzero temperature, new tensor
structures arise in this correlator;
these structures are 
why the Sutherland-Veltman theorem
is inapplicable at $T \neq 0$.  Nevertheless, when
all terms are added together, we find that the Adler-Bardeen
theorem remains valid to one loop order.  
This is a useful and nontrivial
check of our result for $\pi^0 \rightarrow \gamma \gamma$.

Technical details are relegated to several appendices.  The 
WZW action is discussed in appendix~\ref{awzw}.
Various formulas for hard thermal loops are collected in
appendix~\ref{ahtl}.  
We use the imaginary time formalism at nonzero
temperature in this paper, but show 
in appendix~\ref{sec:realtime} how the same
results follow in the real time formalism.
Lastly, in appendix~\ref{gamma3pi} we compute another
anomalous amplitude, that for 
$\gamma \rightarrow \pi \pi \pi$~\cite{bijnens_1},
at low temperature.

While the principal concern of our work are thermal
field theories, we hope that some of our dicussion,
especially that in sec.~\ref{sec:ward},
might be of more general interest.
Perhaps understanding why
$\pi^0 \rightarrow \gamma \gamma$ is not tied 
to the axial anomaly at nonzero temperature helps us
better understand this relation at zero temperature.

\section{$\pi \rightarrow \gamma \gamma$ in vacuum}
\label{sec:vac}

We start by computing the effects of pion loops 
on the amplitude for $\pi \rightarrow \gamma \gamma$ in vacuum.
At one loop order, the relevant diagrams are figs. (1.a)-(1.d):
fig.~(1.a) gives the pion field renormalization constant, $Z_\pi$;
fig.~(1.b), the renormalized pion decay constant $f_\pi$, 
while corrections to the amplitude itself are given by
figs. (1.c) and (1.d).

\begin{figure}[hbt]
\centerline{\epsfig{figure=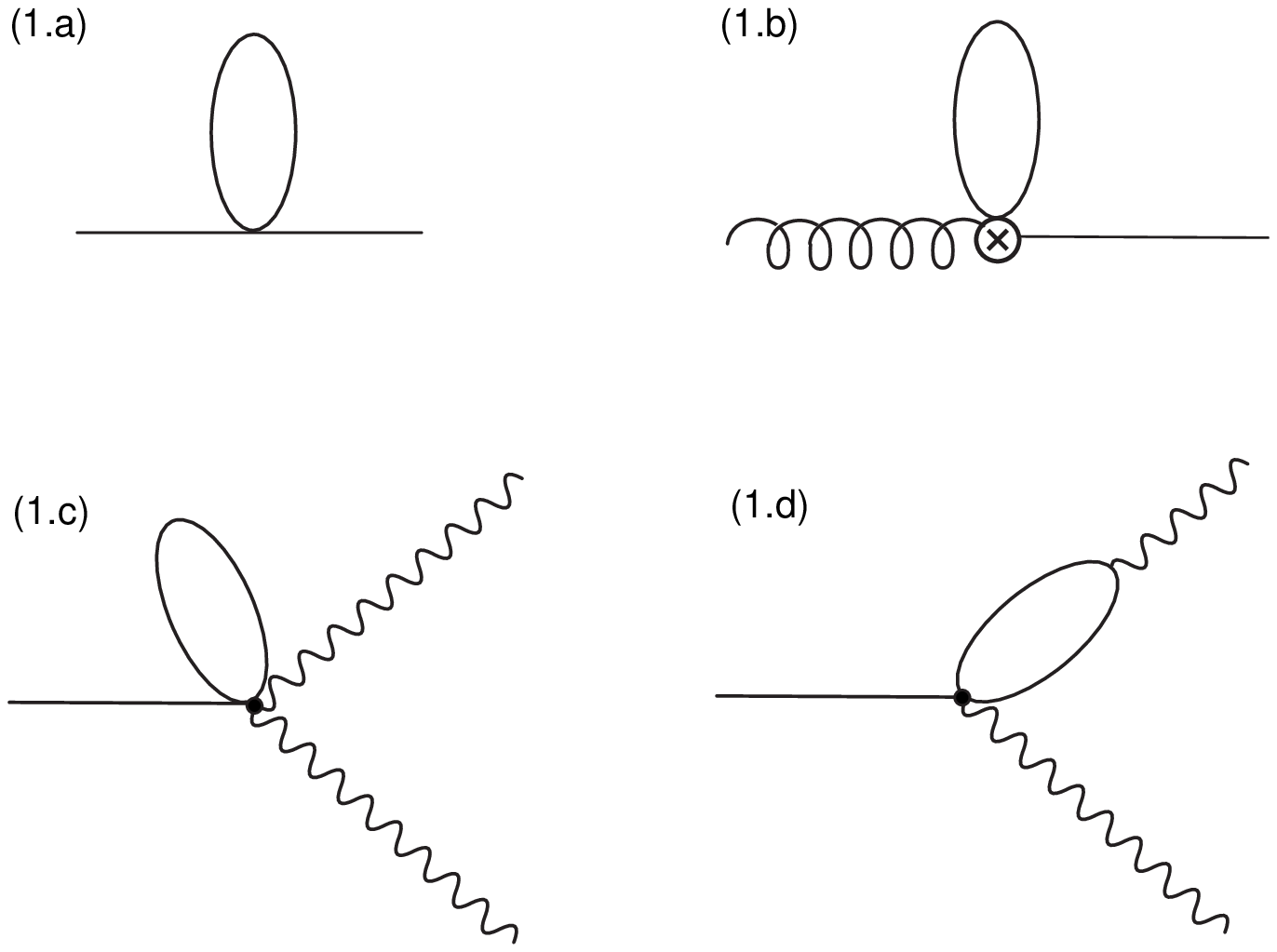,height=50mm}}
\end{figure}
For the ``tadpole'' type diagrams of figs. (1.a), (1.b), and (1.c)
we use a trick.  To compute fig. (1.a) we 
expand the full lagrangian,
(\ref{eq:a.10}), to quartic order in the pion field,
\begin{eqnarray}
\label{eq:2.3}
{\cal L} = {1\over 2} (\partial_\alpha {\vec \pi})^2+ {1\over 6 f_b^2}
[({\vec \pi}\cdot \partial_\alpha {\vec \pi})^2 - {\vec \pi}^2
(\partial_\alpha {\vec \pi})^2] + \ldots 
\end{eqnarray}
In all expressions from the appendix, we need to 
use the bare pion decay constant, $f_b$, instead of
the renormalized quantity $f_\pi$, as one loop effects change
$f_b$ into $f_\pi$.  For the quartic terms, contracting 
two out of the four pion fields in all possible ways gives
\begin{equation}
\label{eq:2.4}
\langle {\cal L} \rangle \simeq {1\over2 } \left ( 1 - {2\over
3}{{\cal I}_0\over f_b^2}\right) (\partial_\alpha {\vec\pi})^2 
\equiv {1 \over2 } (\partial_\alpha {\vec\pi_r})^2  \; .
\end{equation}
where 
\begin{equation}
\label{eq:2.5}
{\cal I}_0 = \langle \pi^2 \rangle = \int
{d^4\!K\over (2 \pi)^4} \; {1\over K^2} \; .
\end{equation}
While this integral is quadratically divergent, 
we ignore regularization, since its actual value is
irrelevant for our purposes.
In (\ref{eq:2.4}) $\pi_r = \pi/\sqrt{Z_\pi}$ 
is the renormalized pion field, and so
\begin{equation}
Z_\pi =  1 + \frac{2}{3} \frac{{\cal I}_0}{f_b^2} \; .
\label{eq:2.5a}
\end{equation}

For the pion decay constant, instead of fig. (1.b)
we expand the axial current
$J_{5,\alpha}^{a}$ of (\ref{eq:a.11}) to 
cubic order in the pion field, 
\begin{equation}
\label{eq:2.6}
J_{5,\alpha}^{a} = f_b \; 
\partial_\alpha \pi^a - {2 \over 3 f_b} \; (\vec \pi^2 \,
\partial_\alpha \pi^a - \pi^a \, \vec \pi \cdot \partial_\alpha 
\vec\pi) + \ldots 
\end{equation}
Contracting all pairs of pion fields,
\begin{equation}
\label{eq:2.7}
\langle J_{5,\alpha}^{a} \rangle =
\left( 1 - {4\over 3} {{\cal I}_0\over f_b^2} \right) \; 
f_b \; \partial_\alpha \pi^a
\equiv f_\pi \partial_\alpha \pi_r^a \; ,
\end{equation}
so that 
\begin{equation}
\label{eq:2.8}
f_\pi = \left ( 1 - {{\cal I}_0\over f_b^2}\right) f_b \; .
\end{equation}  
As is typical of nonlinear sigma models, unphysical, off-shell 
quantities such as $Z_\pi$, (\ref{eq:2.5a}), 
depend upon the parametrization
of the coset space, while physical expressions, such as that 
for $f_\pi$ in (\ref{eq:2.8}), do not.

Turning to the amplitude for 
$\pi^0 \rightarrow \gamma \gamma$, from~(\ref{eq:1.1}) it is
\begin{equation}
\label{eq:2.9}
{\cal M} = g^b_{\pi \gamma \gamma} \,
\varepsilon_{\alpha\beta\gamma\delta}\,
\epsilon_1^\alpha \, \epsilon_2^\beta \,
P_1^\gamma \, P_2^\delta \; ;
\end{equation}
$P_1$ and $P_2$, and $\epsilon_1$ and $\epsilon_2$ 
are the momenta and the polarization vectors of the two photons,
both of which lie on the mass shell,
$P_1^2 = P_2^2 = P_1 \cdot \epsilon_1 
= P_2 \cdot \epsilon_2 = 0$.
At tree level, the bare coupling $g^b_{\pi \gamma \gamma}$ satisfies
\begin{equation}
\label{eq:2.10}
f_b \,g^b_{\pi \gamma \gamma} = {e^2 N_c \over 12 \pi^2} \; .
\end{equation}
The right hand side of (\ref{eq:2.10}) is precisely the coefficient
of the axial anomaly in QED~\cite{adler}.

To evaluate fig. (1.c) we expand the anomalous current
for the coupling to two photons, (\ref{eq:a.7}), (\ref{eq:a.11}), 
and (\ref{eq:a.14}), to cubic order in the pion field,
\begin{eqnarray}
\label{eq:2.11}
{\cal L}_{\pi\gamma\gamma} &=& \left ( {e^2 N_c\over 48
\pi^2}\right )\,{1\over f_b}\, 
\varepsilon_{\alpha\beta\gamma\delta} F_{\alpha\beta} A_\gamma
\times\nonumber\\ 
&&\left[ \left(1 - {2\over 3} {\vec\pi^2\over f_b^2}\right) 
\partial_\delta \pi^3 + {2\over 3 f_b^2} \; 
{\vec\pi}\cdot \partial_\delta {\vec\pi} \; \pi^3 \right] + \ldots 
\end{eqnarray}
Contracting two pion fields,
\begin{equation}
\label{eq:2.12}
\langle {\cal L}_{\pi \gamma\gamma}\rangle
\simeq \left (1 - {4\over 3} {{\cal I}_0\over f_b^2}\right
)\left ( {e^2 N_c\over 48 \pi^2}\right )
\varepsilon_{\alpha\beta\gamma\delta} 
F_{\alpha\beta} A_\gamma \partial_\delta \pi^3 \; .
\end{equation}
The two distinct diagrams of fig. (1.d) require more effort:
\begin{eqnarray}
\label{eq:2.13}
{\cal M}^{d} &=&  8 \left({e^2 N_c \over 48
\pi^2}\right){1\over f_b^3} 
\varepsilon_{\alpha\beta\gamma\delta} 
P^{\beta}_1 \,P^{\delta}_2 \,\epsilon^\alpha_1
\Gamma_{\gamma\sigma}(P_2)\,\epsilon^\sigma_2\nonumber\\
&&\mbox{} +  (P_1, \epsilon_1 \rightleftharpoons\;
P_2, \epsilon_2)  \; ,
\end{eqnarray}
where
\begin{eqnarray}
\label{eq:2.14}
\Gamma^{\alpha\beta}(P) &=& \int {d^4\!K\over (2\pi)^4}\,
 {K^\alpha K^\beta \over K^2 (K-P)^2} \nonumber \\
&\equiv& {{\cal I}_0 \over 2} \, \left( \delta^{\alpha\beta} - 
\Pi^{\alpha\beta}(P) \right) \; .
\end{eqnarray}
This peculiar separation of terms in $\Gamma^{\alpha\beta}(P)$ is
done in anticipation of the results at nonzero temperature.
In vacuum, for $P^2 = 0$,
$\Pi^{\alpha\beta}(P) \sim P^\alpha P^\beta$, and
so because of the Levi-Civita symbol, 
$\Pi^{\alpha\beta}(P)$ does not contribute
to ${\cal M}^{d}$.
Hence (\ref{eq:2.13}) reduces to a form proportional to
the original term in (\ref{eq:2.9}).  Putting everything together,
the renormalized coupling $g_{\pi \gamma \gamma}$ equals
\begin{eqnarray} 
g_{\pi \gamma \gamma} &=& \left(1 +
\left(  {1 \over 3} - {4 \over 3} + 2 \right) 
{{\cal I}_0\over f_b^2} \right) \; 
g^b_{\pi \gamma \gamma} \nonumber \\
&=& \left( 1 + {{\cal I}_0\over f_b^2} \right) 
g^b_{\pi \gamma \gamma} \; .
\label{eq:2.16}
\end{eqnarray}
The $1/3$ comes from a factor of $\sqrt{Z_\pi}$ for the
renormalized pion field, fig. (1.a) and (\ref{eq:2.5a}),
the $-4/3$ from fig. (1.c), (\ref{eq:2.12}), 
and the $2$ from fig. (1.d), (\ref{eq:2.13}) and (\ref{eq:2.14}). 
Hence at one loop order,
\begin{equation}
\label{eq:2.17}
f_\pi \,g_{\pi \gamma \gamma} = 
{e^2 N_c \over 12 \pi^2} \; .
\end{equation}
Comparing (\ref{eq:2.10}) and (\ref{eq:2.17}),
we see that the anomaly is not renormalized
to one loop order~\cite{adler}:
seperate divergences in $f_\pi$ and $g_{\pi \gamma \gamma}$
cancel in the product~\cite{wzwloop}.

\section{$\pi \rightarrow \gamma \gamma$ at low temperature}
\label{sec:lowt}

We now compute the decay for a cool pion,
at a temperature $T \ll f_\pi$~\cite{pistyt2}.
The diagrams are identical, the only difference is that
we need to compute at $T\neq 0$.
For the tadpole diagrams of figs. (1.a), (1.b), and (1.c), the
integral is the analogy of (\ref{eq:2.5}):
\begin{equation}
\label{eq:3.1}
\langle \pi^2 \rangle = 
T \sum_{n = - \infty}^{+ \infty} \int {d^3\!k\over (2\pi)^3}
{1\over K^2 } \; .
\end{equation}
We use the imaginary time formalism,
$K^2 = k_0^2 + \vec{k}^2$, $k = |\vec{k}|$, and
$k^0 = 2 \pi n T$ for integral $n$;
after doing the sum over $n$, the Bose-Einstein statistical distribution
function, $n(\omega) = 1/(\exp(\omega/T) - 1)$, appears:
\begin{eqnarray}
\langle \pi^2 \rangle &=& \int {d^3\! k\over (2 \pi)^3}
\; {1\over 2 k} \; \left(1 + 2 n(k) \right) \nonumber \\
&\equiv& {\cal I}_0 + {\cal I}_T 
= {\cal I}_0 + {T^2 \over 12} \; ;
\label{eq:3.2}
\end{eqnarray}
${\cal I}_0$ is the value of the integral at zero temperature.
Henceforth we drop the $T=0$ part of any integrals,
assuming that they turn bare into renormalized
quantities, such as $f_b$ into $f_\pi$, throughout.

The calculation of temperature dependent corrections
to the pion decay constant proceeds as in the previous
section.  Ignoring ultraviolet renormalization, in (\ref{eq:2.8})
we replace ${\cal I}_0$ by ${\cal I}_T$,
and $f_b$ by $f_\pi$, to obtain
\begin{equation}
f_\pi(T) = \left( 1 - \frac{{\cal I}_T}{f_\pi^2} \right) f_\pi  
= \left( 1 - \frac{1}{12} \frac{T^2}{f_\pi^2} \right) f_\pi \; .
\label{eq:3.2a}
\end{equation}
which was quoted in the introduction, (\ref{eq:1.2}).

Thus if anything unusual happens at nonzero temperature,
it can only be from the diagram of fig. (1.d).
Unlike the tadpole diagrams, this diagram has nontrivial
momentum dependence, and so we must be more precise
in specifying the external momenta.
To compute scattering in a thermal
bath, we continue the euclidean momenta $p^0$
to a minkowski energy $\omega$ by $p^0 = - i \omega + 0^+$.  
Following~\cite{pistyt2} we further assume that each momentum
is not only cool but soft, taking both $|\omega|,p \ll T \ll f_\pi$.

For scattering between
soft, cool pions, in~\cite{pistyt2}
we showed that the leading temperature corrections are
directly analogous to the hard thermal loops of hot
gauge theories~\cite{htl}.  We used
the background field method, but only in the absence of
external gauge fields.  While the perturbative calculations
which follow are thus less elegant, they illustrate
the physics more directly.  From the perspective of \cite{pistyt2},
there is nothing special about 
$\pi^0 \rightarrow \gamma \gamma$;
the connection to the axial anomaly will be clarified later.

We introduce $\delta\Gamma^{\alpha\beta}(P)$, 
\begin{equation}
\delta \Gamma^{\alpha\beta}(P) 
= \frac{T^2}{24} \left( \delta^{\alpha\beta}
- \delta\Pi^{\alpha\beta}(P) \right)\; ,
\label{eq:3.3}
\end{equation}
with
\begin{equation}
\frac{T^2}{24} \; \delta\Pi^{\alpha\beta}(P) 
\approx
T \sum_{n = - \infty}^{+ \infty} \int {d^3\!k\over (2\pi)^3}
\left\{ { \delta^{\alpha\beta} \over 2 } 
{1\over K^2} 
 - {K^\alpha K^\beta \over K^2 (K-P)^2} \right\} \; .
\label{eq:3.4}
\end{equation}
The $\approx$ sign denotes that only the hard thermal loops
in the integral are retained, which we denote by
$\delta \Gamma^{\alpha \beta}(P)$ and 
$\delta\Pi^{\alpha\beta}(P)$.
The hard thermal loops are the terms $\sim T^2$,
and are given explicitly in appendix \ref{ahtl}.

Up to an overall constant, $\delta \Pi^{\alpha \beta}$
is the same hard thermal loop as appears in the polarization
tensor for a photon in thermal equilibrium.
For a thermal photon, the screening of time 
dependent electric and magnetic fields implies that
the mass shell is at $P^2 \sim e^2 T^2$.
For the sake of simplicity we assume 
that the photons do not thermalize;
then the only photons which propagate are
transverse modes on the light cone,
$P_1^2 = P_2^2 = 0$.  The polarization vectors
for these modes are purely spatial
vectors which satisfy $P \cdot \epsilon = 0$.

From (\ref{eq:2.13}), the contribution to the amplitude
from fig.~(1.d), ${\cal M}^{d}$, involves
$\Gamma^{\alpha \beta}(P) \epsilon^\beta$
for one of the two photons on their mass shell.  
Using (\ref{eq:b.5}) of appendix~\ref{ahtl}, 
\begin{equation}
\label{eq:3.6}
\delta \Pi^{\alpha \beta}(P) \; \epsilon^\beta|_{P^2 = 0}
= \epsilon^\alpha \; ,
\end{equation}
where $\epsilon^\beta$ is the polarization vector for the photon
with momentum $P$.
Only the first term on the right hand side of (\ref{eq:b.5})
contributes, as terms in 
$\delta \Pi^{\alpha \beta}(P)$ which are 
$\sim P^\beta$ or $n^\beta$ 
drop out after contraction with
$\epsilon^{\beta}$.
From the definition of $\delta \Gamma^{\alpha \beta}$,
(\ref{eq:3.3}), 
\begin{equation}
\delta \Gamma^{\alpha \beta}(P) \,
\epsilon^\beta|_{P^2=0} = 0 \; .
\label{eq:3.7}
\end{equation}
Consequently, while at zero temperature
fig.~(1.d) contributes to the amplitude for
$\pi^0 \rightarrow \gamma \gamma$,
to leading order at nonzero temperature
its contribution vanishes identically,
(\ref{eq:3.7}).

Knowing that fig.~(1.d) doesn't contribute,
it is then easy to read off 
the one loop corrections to the coupling $g_{\pi \gamma \gamma}$ 
to $\sim T^2/f_\pi^2$, $g_{\pi \gamma \gamma}(T)$.
As in (\ref{eq:3.2a}), we start with
(\ref{eq:2.16}), and then replace ${\cal I}_0$ by ${\cal I}_T$, and
$f_b$ by $f_\pi$.  We keep
the $1/3$ from fig.~(1.a), the 
the $-4/3$ from fig.~(1.c), but replace the $+2$
from fig.~(1.d) by $0$, to obtain
\begin{eqnarray}
g_{\pi \gamma \gamma}(T) 
&=& \left(1 + \left(\frac{1}{3} - \frac{4}{3} + 0 \right)
\frac{{\cal I}_T}{f_\pi^2} \right) g_{\pi \gamma \gamma} \nonumber \\
&=& \left(1 - {{\cal I}_T \over f_\pi^2}\right) 
\; g_{\pi \gamma \gamma} \; .
\label{eq:3.8}
\end{eqnarray}
Notice that while
(\ref{eq:2.8}) is precisely analogous to (\ref{eq:3.2a}),
because of the difference in fig.~(1.d), 
(\ref{eq:2.16}) is not analogous to (\ref{eq:3.8}) --- there
is a difference in sign.  Consequently,
\begin{equation}
g_{\pi \gamma \gamma}(T) =
\left(1 - {1 \over 12} \, {T^2 \over f_\pi^2}\right) 
\; g_{\pi \gamma \gamma} \; . 
\label{eq:3.8a}
\end{equation}
As discussed in the introduction, naively one might guess that
(\ref{eq:2.17}) generalizes to nonzero temperature 
just by replacing $f_\pi$ and $g_{\pi \gamma \gamma}$ with
$f_\pi(T)$ and $g_{\pi \gamma \gamma}(T)$, respectively.
This is wrong: to $\sim T^2/f_\pi^2$,
instead of $g_{\pi \gamma \gamma}(T) \sim 1/f_\pi(T)$, 
as would be guessed from (\ref{eq:2.17}), instead
$g_{\pi \gamma \gamma}(T) \sim f_\pi(T)$, (\ref{eq:3.8a}).
We do not know why, to leading order about low temperature, 
$g_{\pi \gamma \gamma}(T)$ decreases in exactly the
same manner as $f_\pi(T)$.

Our result in (\ref{eq:3.8}) differs from 
that found by Dobado, Alvarez-Estrada, and Gomez~\cite{estrada2}.
These authors consider the same model, but find
$g_{\pi \gamma \gamma}(T) = g_{\pi \gamma \gamma}$ 
to $\sim T^2/f_\pi^2$.  Our results agree except for 
fig.~(1.d), which we believe was treated incorrectly~\cite{estrada3}.

Before continuing, following \cite{pistyt2}
we construct the effective lagrangian
for $\pi^0 \rightarrow \gamma \gamma$ 
to $\sim T^2/f_\pi^2$.
In $\delta \Gamma^{\alpha\beta}(P)$ of 
(\ref{eq:3.3}), the term $\sim \delta^{\alpha\beta}$
is easy to include.  At zero temperature, (\ref{eq:2.14}),
this term is the only part of fig.~(1.d) which contributes,
$+2$ in (\ref{eq:2.16}), and turns $1/f_b$ in $g_{\pi \gamma \gamma}$
into $1/f_\pi$ in $g_{\pi \gamma \gamma}^r$.
Thus to $\sim T^2/f_\pi^2$, the effect
of figs. (1.a), (1.c), and the term $\sim \delta^{\alpha\beta}$
in $\delta \Gamma^{\alpha\beta}(P)$ of fig.~(1.d)
is just to change $1/f_\pi$ into $1/f_\pi(T)$
in the original lagrangian, 
${\cal L}_{\pi\gamma\gamma}$ of (\ref{eq:1.1}).

Including the term $\delta\Pi^{\alpha\beta}(P)$ is less trivial.
Because of (\ref{eq:b.1}), it must be
constructed out of transverse quantities.  Using
(\ref{eq:b.7}),
we find that to $\sim T^2/f_\pi^2$,
the effective lagrangian for $\pi^0 \rightarrow \gamma \gamma$ is
\begin{equation}
\label{eq:3.9}
{\cal L}_{\pi^0 \gamma \gamma}(T) 
=  \left( {e^2 N_c\over 48 \pi^2} \right)  \; {1 \over f_\pi(T)}\;
\pi^0 F_{\alpha \beta} \widetilde{F}^{\alpha \beta}
\end{equation}
$$
 - \frac{T^2}{12 f_\pi^2} \left( {e^2 N_c\over 48 \pi^2} \right)
\int \frac{d \Omega_{ \hat{k} } }{4 \pi} \; 
H_{\gamma \alpha}
\frac{ \hat{K}^\alpha \hat{K}^\beta}{- (\partial \cdot \hat{K})^2}
F_{\gamma \beta} \; .
$$
In this expression $\widetilde{F}^{\alpha \beta} =
\epsilon^{\alpha \beta \gamma \delta} F_{\gamma \delta}/2$,
\begin{equation}
H_{\alpha\beta} = \partial_\alpha H_\beta - \partial_\alpha H_\beta \; ,
\label{eq:3.10}
\end{equation}
and
\begin{equation}
H_\alpha = {1\over f_\pi}\varepsilon_{\alpha\beta\gamma\delta} 
F_{\beta\gamma} \partial_\delta \pi^0 \; .
\label{eq:3.11}
\end{equation}
The vector $\hat K = ( i, \hat{k})$ and the integration
over the angle $\hat{k}$ arediscussed following (\ref{eq:b.2}).

The non-local term in~(\ref{eq:3.9}) is specific to finite
temperature.  At zero temperature, there is no other term
besides~(\ref{eq:1.1})
that  contributes to  $\pi^0\rightarrow
\gamma\gamma$ for photons on the mass-shell.
In the terminology of the nonlinear sigma model \cite{chiral},
the operator of~(\ref{eq:1.1}) is ${\cal O}(P^4)$, 
while operators at next to leading order are ${\cal O}(P^6)$.
These operators, however, are either 
proportional to $P_1^2$ or $P_2^2$, and so vanish on 
the photon(s) mass shell, or $m^2_\pi$, and so vanish in the
chiral limit.  Thus in the
vacuum, the only possible change in (\ref{eq:1.1})
is the transmutation of $f_b$ into $f_\pi$, sec. \ref{sec:vac}.
At nonzero temperature, however, there are new nonlocal terms
which arise,~(\ref{eq:3.9}).  Because they are nonlocal,
these new terms are also ${\cal O}(P^4)$, and so as important
as (\ref{eq:1.1})\cite{pistyt2}.  This is the technical reason why the
amplitude for $\pi^0\rightarrow \gamma\gamma$ 
depends nontrivially upon temperature.

\section{$\pi^0 \rightarrow \gamma\gamma$ and the axial anomaly}
\label{sec:ward}

In the previous section we found that the amplitude for
$\pi^0 \rightarrow \gamma\gamma$
diminishes to leading order in an expansion about low
temperature, (\ref{eq:3.8a}).  The question we address
in this section is why is this amplitude tied to
the coefficient of the axial anomaly at zero temperature,
(\ref{eq:2.17}), but not at nonzero temperature?

We work in the chiral limit
to leading order about zero temperature, $\sim T^2/f_\pi^2$,
because then we can make certain technical assumptions which
simplify the discussion.  The general case is considered at
the end of this section.

Define the vector current,
$J_\alpha$, and
the axial current in the isospin-$3$ direction, $J_{5,\gamma}^3$.
The vector curent is conserved, 
\begin{equation}
\partial^\alpha J_\alpha = 0 \; ,
\end{equation}
while the axial current is anomalous,
\begin{equation}
\partial^\alpha J_{5,\alpha}^3
= - \frac{e^2 N_c}{48 \pi^2} F_{\alpha \beta} 
\widetilde{F}^{\alpha \beta} \; .
\label{eq:4.0a}
\end{equation}
By the Adler-Bardeen theorem,
the coefficient of the right hand side is
exact to one loop order~\cite{adler}, 
and is independent of temperature and
density~\cite{anom2,mueller,anom4}.

One quantity we can compute is the
(thermal) three point Green's function between
two vector, and one axial vector, current:
\begin{eqnarray}
\label{eq:4.1}
{\cal T}_{\alpha\beta\gamma}(P_1,P_2;T) 
&=& - i \, e^2 \int d^4 X_1 d^4 X_2\, 
e^{i(P_1\cdot X_1 + P_2\cdot X_2)}\,\\
& \times & 
\frac{\mbox{\rm Tr}\left(e^{-H/T} J_\alpha(X_1) 
J_\beta(X_2) J^3_{5,\gamma}(0) \right)}{\mbox{\rm Tr}(e^{-H/T})}
\; , \nonumber
\end{eqnarray}
where $H$ is the hamiltonian.  Then
${\cal T}_{\alpha \beta \gamma}$ satisfies current conservation,
\begin{equation}
\label{eq:4.3}
P_1^\alpha {\cal T}_{\alpha\beta\gamma} 
= P_2^\beta  {\cal T}_{\alpha\beta\gamma} = 0 \; ,
\end{equation}
and the anomalous Ward identity,
\begin{equation}
\label{eq:4.2}
Q^\gamma {\cal T}_{\alpha\beta\gamma} = - {e^2 N_c\over 12 \pi^2}
\,\varepsilon_{\alpha\beta\gamma\delta}\, P_1^\gamma P_2^\delta \; ,
\end{equation}
$Q = P_1 + P_2$.  

To relate the anomalous Ward identity to the amplitude
for pion decay
we follow Shore and Veneziano~\cite{shore}. 
At low temperature the pion couples to the axial current as
\begin{equation}
\langle 0\vert J_{5, \alpha}^a\vert \pi^b(Q)\rangle 
= i Q_\alpha f_\pi \delta^{a b} \; .
\label{eq:4.4}
\end{equation}
(This is not valid to $\sim T^4/f_\pi^4$; then the relation
is more complicated, \cite{pistyt1}.)

To obtain the amplitude for $\pi^0\rightarrow \gamma \gamma$,
we introduce $Q^2$ times
the matrix element between two $QED$ currents
and a pion, 
\begin{eqnarray}
\label{eq:4.6}
{\cal T}_{\alpha\beta} &=&  e^2 Q^2 \int d^4 X_1 d^4 X_2\, 
e^{i(P_1\cdot X_1 + P_2\cdot X_2)}\\
&\times & 
\frac{\mbox{\rm Tr}\left(e^{-H/T} J_\alpha(X_1) 
J_\beta(X_2) \pi(0)\right)}{\mbox{\rm Tr}\left(e^{-H/T}\right)}
\; . \nonumber
\end{eqnarray}
This is related to the pion decay amplitude, (\ref{eq:2.9}), as
\begin{equation}
\label{eq:4.7}
{\cal M} = \lim_{Q^2\rightarrow 0}
\epsilon_1^\alpha \epsilon_2^\beta \,
{\cal T}_{\alpha\beta} \; .
\end{equation}

Subtracting the one pion pole term from (\ref{eq:4.1})
gives $\widehat {\cal T}_{\alpha\beta\gamma}$,
which by construction is one pion irreducible,
\begin{equation}
\label{eq:4.5}
\widehat {\cal T}_{\alpha\beta\gamma} = {\cal T}_{\alpha\beta\gamma} +  
f_\pi  \; Q_\gamma \,  { 1\over Q^2} \,
{\cal T}_{\alpha\beta} \; .
\end{equation}
Again, by current conservation
\begin{equation}
\label{eq:4.9}
P_1^\alpha \widehat {\cal T}_{\alpha\beta\gamma} = P_2^\beta
\widehat {\cal T}_{\alpha\beta\gamma} = 0 \; ,
\end{equation}
while the anomalous Ward identity, (\ref{eq:4.2}), becomes
\begin{eqnarray}
\label{eq:4.8}
Q^\gamma \widehat {\cal T}_{\alpha \beta \gamma} =  f_\pi \;
{\cal T}_{\alpha\beta} - { e^2 N_c\over 12 \pi^2}
\,\varepsilon_{\alpha\beta\gamma\delta}\, P_1^\gamma P_2^\delta \; .
\end{eqnarray}
The trick is now to try to deduce general relations using
just (\ref{eq:4.9}), (\ref{eq:4.8}), and Bose symmetry between
the two photons,
$P_1, \alpha \rightleftharpoons P_2, \beta$.

We first discuss
zero temperature, where we can invoke euclidean invariance.
The most general pseudo-tensor $\widehat {\cal T}_{\alpha\beta\gamma}$ 
which satisfies all of our conditions involves three terms:
\begin{eqnarray}
\label{eq:4.10}
\widehat{\cal T}_{\alpha\beta\gamma} &=& T_1\, 
\varepsilon_{\alpha\beta\gamma\delta}( P_1^\delta
- P_2^\delta) \\
&+& T_2 \,(\varepsilon_{\alpha\gamma \delta \kappa} P_{2}^\beta\, - 
\varepsilon_{\beta\gamma \delta \kappa} P_1^\alpha) P_1^\delta
P_2^\kappa \nonumber \\
&+&  T_3\, (\varepsilon_{\alpha\gamma \delta \kappa} P_1^\beta - 
\varepsilon_{\beta\gamma \delta \kappa} P_2^\alpha) 
P_1^\delta P_2^\kappa \; . \nonumber 
\end{eqnarray}
Current conservation, (\ref{eq:4.9}), gives
\begin{equation}
\label{eq:4.12}
T_1 + P_1^2 \, T_2 + P_1\cdot P_2 \, T_3 = 0 \; ,
\end{equation}
while from the anomalous Ward identity, (\ref{eq:4.8}), 
\begin{equation}
\label{eq:4.11}
- 2\, T_1 = f_\pi g_{\pi\gamma\gamma} - {e^2 N_c\over 12 \pi^2} \; .
\end{equation}
Combining these two relations we obtain
\begin{equation}
\label{eq:4.13}
2 P_1^2 \, T_2 + 2 P_1 \cdot P_2  \, T_3
= f_\pi g_{\pi\gamma\gamma} - {e^2 N_c\over 12 \pi^2} \; .
\end{equation}
Putting the photons on their mass shell $P_1^2 = P_2^2$, 
the left hand side in~(\ref{eq:4.13}) reduces to $Q^2 T_3$.
This is zero on the pion mass shell, $Q^2 \rightarrow 0$,
since by definition $\widehat{\cal T}$ is one pion irreducible,
and so cannot have a pole $\sim 1/Q^2$.
Hence the left hand side of (\ref{eq:4.13}) vanishes, and
we obtain the desired relation between $g_{\pi \gamma \gamma}$
and the coefficient of the axial anomaly,
\begin{equation}
\label{eq:4.14}
0 = f_\pi g_{\pi\gamma\gamma} - {e^2 N_c\over 12 \pi^2} \; ,
\end{equation}
which is~(\ref{eq:2.14}).

This analysis, and especially the tensor
decomposition of (\ref{eq:4.10}), is identical to the 
derivation of the
Sutherland-Veltman theorem~\cite{suther-veltman}. 
Historically, this theorem predated the anomaly, 
and was used to conclude that $g_{\pi \gamma \gamma} = 0$.  
By adding the axial 
anomaly through the anomalous Ward identity of 
(\ref{eq:4.8}), however, we obtain 
$g_{\pi \gamma \gamma} \sim e^2 N_c/f_\pi$, (\ref{eq:4.14}).
From a modern perspective, then, it is precisely the 
Sutherland-Veltman theorem which relates
the axial anomaly
to the amplitude for $\pi^0 \rightarrow \gamma \gamma$,
and tells us at zero temperature
the left hand side of (\ref{eq:4.14}) vanishes.

This is no longer true at nonzero temperature.  
Following Itoyama and Mueller~\cite{mueller},
we write the most general tensor decomposition
of $\widehat {\cal T}_{\alpha \beta \gamma}$.  In
a thermal bath, however, euclidean symmetry is lost,
and the rest frame of the thermal 
bath, which we take as $n_\mu = (1,\vec{0})$, enters.
Some of the possible tensors include
\begin{eqnarray}
\label{eq:4.15}
\widehat{\cal T}_{\alpha\beta\gamma} &=& T_1\, 
\varepsilon_{\alpha\beta\gamma\delta}( P_1^\delta - P_2^\delta) \\
&& \mbox{} + T_2 \,(\varepsilon_{\alpha\gamma\delta\kappa} \, P_2^\beta - 
\varepsilon_{\beta\gamma\delta\kappa}\, P_1^\alpha) P_1^\delta
P_2^\kappa \nonumber \\
&& \mbox{} + T_3\, (\varepsilon_{\alpha\gamma\delta\kappa} \, P_1^\beta - 
\varepsilon_{\beta\gamma\delta\kappa}\, P_2^\alpha ) P_1^\delta P_2^\kappa
\nonumber\\
&& \mbox{} + T_4 \, n\cdot Q \, 
\varepsilon_{\alpha\beta\delta\kappa} \, P_1^\delta P_2^\kappa \,
n^\gamma\nonumber\\
& & \mbox{} + T_5 
(n\cdot P_2 \, \varepsilon_{\alpha\gamma\delta\kappa} \, n^\beta  -
n\cdot P_1 \, \varepsilon_{\beta\gamma\delta\kappa} \, n^\alpha ) 
P_1^\delta P_2^\kappa\nonumber
\\
&& \mbox{} + \ldots\nonumber
\end{eqnarray}
We have only included the terms in $\widehat{\cal T}$ which
contribute to the Ward identities of (\ref{eq:4.9}) and 
(\ref{eq:4.8}).  Current conservation gives
\begin{equation}
\label{eq:4.17}
T_1 + P_1^2 \;T_2 + P_1 \cdot P_2\; T_3 
+ (n\cdot P_1)^2 \; T_5 = 0 \; ,
\end{equation}
while the anomalous Ward identity fixes

\begin{equation}
\label{eq:4.16}
- 2 T_1 + (n\cdot Q)^2 \, T_4  =  f_\pi(T) g_{\pi\gamma\gamma}(T) -  
{e^2 N_c\over 12 \pi^2} \; ,
\end{equation}
from which follows
$$
2 P_1^2 \, T_2 + 2 P_1 \cdot P_2  \, T_3
+(n\cdot Q)^2 \, T_4 + 2 (n \cdot P_1)^2 \, T_5
$$
\begin{equation}
= f_\pi(T) \, g_{\pi\gamma\gamma}(T) -  {e^2 N_c\over 12 \pi^2} \; .
\label{eq:4.18a}
\end{equation}
Putting all fields on their mass shell,
$P_1^2 = P_2^2 = 0$, $Q^2 \rightarrow 0$, and
assuming as before that $T_3$ has no pole 
$\sim 1/Q^2$, we obtain
\begin{equation}
\label{eq:4.18}
(n\cdot Q)^2\, T_4 + 2 (n\cdot P_1)^2\, T_5   
= f_\pi(T) \, g_{\pi\gamma\gamma}(T) -  {e^2 N_c\over 12 \pi^2} \; .
\end{equation}
Since the terms on the left hand side involve only
$n\cdot Q$ and $n \cdot P_1$,
there is no reason for them to vanish even if
$P_1^2=P_2^2=Q^2 =0$; thus the direct connection
between $g_{\pi \gamma \gamma}(T)$, $f_\pi(T)$,
and the anomaly is lost.  We stress that,
as always, the Adler-Bardeen theorem remains
valid, and gives
(\ref{eq:4.18a}).  It is only the Sutherland-Veltman
theorem which no longer applies at nonzero temperature.

The above analysis only applies to
leading order in low temperature, $\sim T^2/f_\pi^2$.
This is because beyond leading order, the pion mass
shell is no longer at $Q^2 = 0$ \cite{pistyt1}; also,
as photons thermalize, their mass shell moves off
the light cone.  This is incorporated by using (\ref{eq:4.18a})
instead of (\ref{eq:4.18}).
For instance, we can understand
how the anomalous Ward identity is satisfied in
a chirally symmetric phase.  From explicit calculation 
in a constituent quark model \cite{rob_1},
$\pi^0 \rightarrow \gamma \gamma$ vanishes 
once chiral symmetry is restored.  This does
not conflict with the anomalous 
Ward identity since even
if $g_{\pi \gamma \gamma}(T)=0$, there are
other terms which can ensure that 
(\ref{eq:4.18a}) is satisfied.  For photons which do
not thermalize, so $P_1^2 = P_2^2 = 0$, even
assuming that at the chiral critical point that
the pion mass shell is $Q^2 = 0$, 
the tensors $T_4$ and/or $T_5$ will
in general be nonzero.

Our analysis agrees with the results of
Contreras and Loewe~\cite{piot}, who computed
the triangle diagram at $T\neq 0$ with massive fermions. 
Their result, (1.2), is 
directly related to $\widehat {\cal T}_{\alpha\beta\gamma}$.
There are two terms: the first is regular,
temperature dependent, and gives the amplitude for
$\pi^0\rightarrow \gamma\gamma$, while
the second is the anomaly, and is independent of
temperature.  

At first sight it might appear peculiar that the
Sutherland-Veltman theorem applies at zero temperature,
but fails at any nonzero temperature.  
Even at zero temperature, however, the Sutherland-Veltman
theorem only applies in the chiral 
limit when both photons are on their mass shell.
Without all of these conditions, the left
hand side of (4.13) does not vanish, and $g_{\pi \gamma \gamma}$
is not given by (4.14).  An example of this occurs
when one (or both) of the photons are off the mass shell \cite{bill},
even if $Q^2 \rightarrow 0$.
In particular, in the limit of large $P^2$, it is known that
$g_{\pi \gamma \gamma} \sim e^2 f_\pi/P^2$ \cite{bill}.  Thus
on the right hand side of (4.13), we can neglect 
$f_\pi g_{\pi \gamma \gamma} \sim 1/P^2$ relative to
the anomaly term, $-e^2 N_c/(12 \pi^2)$, and (4.13)
tells us that a combination of $T_2$ and $T_3$ are
given entirely by the anomaly, $\sim e^2 N_c/P^2$.

\section{The Adler-Bardeen theorem at low temperature}
\label{sec:adlerbardeen}

In this section we calculate  the correlator
$\widehat{\cal T}_{\alpha\beta\gamma}$
to $\sim T^2/f_\pi^2$ in the nonlinear sigma model.
This allows us to check explicitly the temperature dependence
of $f_\pi(T)$ and $g_{\pi \gamma \gamma}(T)$ found
previously in sec. ~\ref{sec:lowt}.  
It is also illuminating to see exactly which amplitudes enter
at one loop order in the nonlinear sigma model, and 
how they conspire to satisfy the Adler-Bardeen theorem.

At tree level in the nonlinear sigma model, 
there is no direct coupling between two vector
and one axial vector currents, so $T_1 = T_2 = T_3 = 0$.
Thus the anomalous Ward identity, (\ref{eq:4.2}),
is satisfied entirely by the one pion reducible term,
(\ref{eq:4.8}), with $\widehat{\cal T}_{\alpha \beta \delta} = 0$.  This is 
illustrated in fig.~(2).

\begin{figure}[hbt]
\centerline{\epsfig{figure=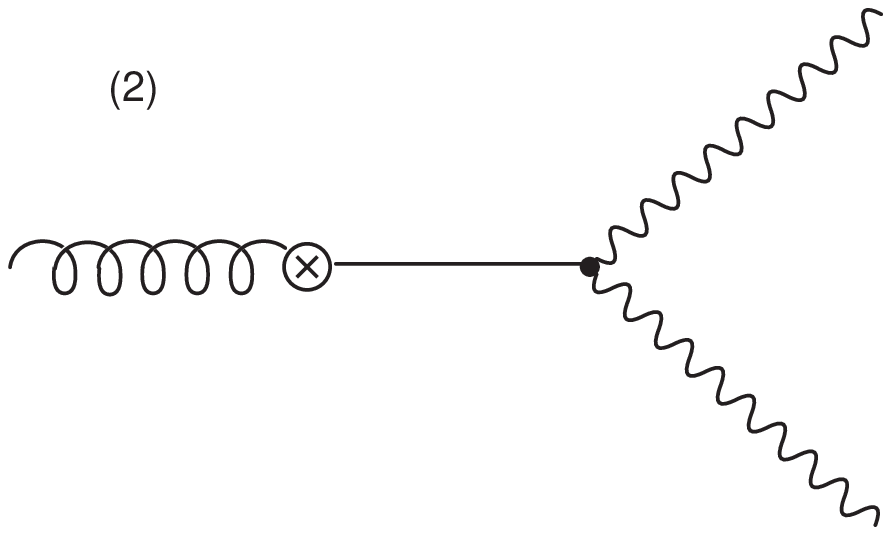,height=25mm}}
\end{figure}

Contributions to $\widehat{\cal T}_{\alpha \beta \gamma}$ 
are generated at one loop order.
In order to compute these, it is necessary to compute corrections
to the axial current.  We have computed these in two ways.
The most direct is to follow the original method of 
Wess and Zumino \cite{wess-zumino}.  
Besides the the photon field $A_\alpha$, 
which couples to the electromagnetic current
$J_\alpha$, we also introduce an external field
$A_{5,\alpha}^3$ which couples to the axial current $J_{5,\alpha}^3$.
One then differentiates the generating
functional of Wess and Zumino with respect to the external fields
$A_\alpha$ and $A_{5,\alpha}^3$.
At one loop order, the terms required are
$\sim \pi^{+} \pi^{-} A_{5,\alpha}^{3} A_\beta$ and
$\sim \pi^{+} \pi^{-} A_{5,\alpha}^{3} A_\beta A_\gamma$;
$\pi^{\pm} = (\pi^1 \pm i \pi^2)/\sqrt{2}$ are the charged
pion fields.
After lengthy calculation, we find
$$
\Delta {\cal L} = \left(
\frac{e N_c}{48 \pi^2 f_\pi^2} \right)
\epsilon_{\alpha \beta \gamma \delta} 
F_{\alpha \beta} A^3_{5,\gamma} \times
$$
\begin{equation}
\left( i 
(\pi^{-}\partial_{\delta}\pi^{+} - \pi^{+}\partial_{\delta}\pi^{-})
- 2 e \pi^+ \pi^- A_\delta \right)
\label{eq:5.1}
\end{equation}
To leading order in $e$, this ${\cal L}$ is 
invariant under an electromagnetic gauge transformation,
\begin{eqnarray}
\pi^{\pm}(x) \rightarrow \exp^{\pm ie \theta(x)}\pi^{\pm}(x)
\nonumber \\ A_{\alpha}(x) \rightarrow  A_{\alpha}(x) - \partial_{\alpha}
\theta(x).
\end{eqnarray}
The terms in ${\cal L}$ can be viewed as corrections
to the axial current.  
In appendix A we compute these corrections using
the Noether construction of the axial current.
This is somewhat delicate, since it is necessary to start with a Lagrangian
density (and not merely a Lagrangian) which is manifestly
gauge invariant.  The result, (\ref{eq:a.19}), agrees
with (\ref{eq:5.1}).

Through the diagrams of fig.~(3), the couplings in
$\Delta{\cal L}$ generate the one pion irreducible terms in
$\widehat {\cal T}_{\alpha \beta \gamma}$.
\begin{figure}[hbt]
\centerline{\epsfig{figure=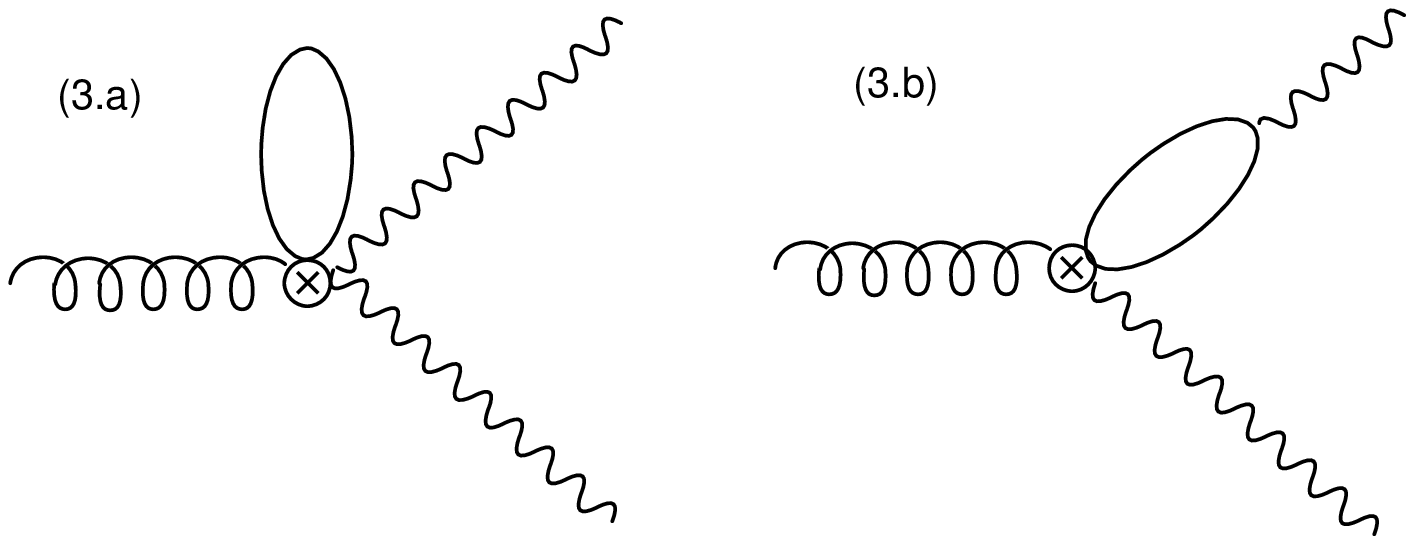,height=25mm}}
\end{figure}

We begin with the results at zero temperature.  
Given the couplings in $\Delta {\cal L}$, the only tensor
structure which arises from fig.~(3) is 
$\sim \epsilon_{\alpha \beta \gamma \delta} (P_1 - P_2)^\delta$.
Comparing with the tensors in~(\ref{eq:4.10}), then,
at one loop order automatically $T_2 = T_3 = 0$.
For $T_1$, we find
\begin{equation}
T_1 = (2 - 2) \left( \frac{{\cal I}_0 }{f_b^2} \right)
\frac{e^2 N_c}{24 \pi^2} 
= 0 \; .
\label{eq:5.1a}
\end{equation}
The contribution $\sim + 2$ is from fig.~(3.a), that
$\sim -2$ from the two diagrams of fig.~(3.b).  
Because all of the $T_i$'s vanish, 
 current conservation~(\ref{eq:4.12})
and the anomaly equation~(\ref{eq:4.11}), are satisfied rather trivially.
That the latter vanishes
is equivalent to the Sutherland-Veltman
theorem,~(\ref{eq:4.14}).  This explains our
results in sec.~\ref{sec:vac}, where we found that at
zero temperature, the renormalization 
of $f_\pi$ and $g_{\pi \gamma \gamma}$ exactly compensate each other.

At nonzero temperature we can evaluate $T_1 \ldots T_5$ using
the results of sec.~\ref{sec:lowt}.
Fig.~(3.a) is a tadpole diagram, and so as at zero temperature, just
contributes to $T_1$.  Thus the only possibility
for a new tensor structure is from the hard thermal
loop in fig.~(3.b).
For one ordering of momenta, fig.~(3.b) gives
\begin{equation}
- i \frac{e^2 N_c}{12 \pi^2 f^2_\pi} 
\epsilon_{\gamma \delta \beta \kappa} P_2^\delta 
\delta \Gamma^{\kappa \alpha}(P_1)
\; .
\label{eq:5.12}
\end{equation}
where $\delta \Gamma^{\kappa \alpha}$ is given in~(\ref{eq:3.3}).  
Putting the photon momentum $P_1$
on its mass shell, we require the result for
$\delta \Pi^{\kappa \alpha}(P_1)$ in~(\ref{eq:b.5}).
We drop all terms in $\delta \Pi^{\kappa \alpha}(P_1) \sim P_1^\alpha$,
(\ref{eq:b.5}), since they vanish
upon contraction with the photon polarization tensor.
The term $\sim \delta^{\kappa \alpha }$ in~(\ref{eq:b.5})
cancels against the same term in~(\ref{eq:3.3});
thus at nonzero temperature, there is no contribution to
$T_1$ from fig.~(3.b), and
\begin{equation}
T_1 = ( 2 + 0) \, \left( {{\cal I}_T\over f_\pi^2}\right)
{e^2 N_c \over 24 \pi^2} \; .
\end{equation}
Comparing to (\ref{eq:5.1a}), again fig.~(3.a) gives the
term $\sim +2$, and fig.~(3.b) the term $\sim 0$.  That
fig.~(3.b) doesn't contribute is just like the same
result for $g_{\pi \gamma \gamma}(T)$, fig.~(1.d) and 
(\ref{eq:3.8}).

Fig.~(3.b) will contribute, however, through
the term $\sim n^\alpha P_1^\kappa/(n \cdot P_1)$ in
$\delta \Pi^{\kappa \alpha}(P_1)$ in~(\ref{eq:b.5}).
Comparing with the tensor decomposition of~(\ref{eq:4.15}),
we find:
\begin{equation}
T_5 =  - { 1\over (n \cdot P_1 )^2} 
\left(\frac{{\cal I}_T}{f_\pi^2} \right) 
\frac{e^2 N_c}{12 \pi^2}  \; .
\end{equation}
Since these are the only diagrams at one loop order, 
\begin{equation}
T_2 = T_3 = T_4 = 0 \; .
\end{equation}
Consequently, 
\begin{equation}
- 2 T_1 =  2 (n \cdot P_1 )^2 \, T_5 =  
f_\pi(T) g_{\pi \gamma \gamma}(T) - \frac{e^2 N_c}{12 \pi^2}.
\label{eq:5.6}
\end{equation}
Thus we can see that our results satisfy current conservation, 
(\ref{eq:4.17}) and the anomaly
equation of (\ref{eq:4.18}).  Last but not least,
the latter only holds given $f_\pi(T)$
and $g_{\pi \gamma \gamma} (T)$ in (\ref{eq:1.1})
and (\ref{eq:3.8}), and so provides a nontrivial check of these
calculations.

Notice that while there is a factor of
$1/(n\cdot P_1)^2$ in
$T_5$, it is essentially kinematic in origin, as envisioned by
Itoyama and Mueller \cite{mueller}. One factor of $n\cdot P_1$
arises from the definition of $T_5$, (\ref{eq:4.15}), while the
other can be seen to arise from a directional singularity, $\sim
p^i/p$, in the integral of~(\ref{eq:b.5}). 
Further, notice that there is a (logarithmic)
collinear divergence in the integral of ~(\ref{eq:b.5}).
This singularity drops out of the full amplitude after
contracting with the polarization tensor of the photon.

\section{Conclusions}

In this paper we have concentrated exclusively on the anomalous
decay of $\pi^0 \rightarrow \gamma \gamma$.  We have done
so because it is the most familiar anomalous decay, and
because the connection to the axial anomaly is especially
close.  For the collisions of heavy ions at ultrarelativistic
energies, though, this is a purely academic point, since any
high temperature plasma flies apart long before a $\pi^0$
has a chance to decay electromagnetically.

Our basic point is much more general, however.  While
the axial anomaly for fermions is
independent of temperature, anomalous decays of mesonic
fields change with temperature.  Processes
of obvious interest for hadronic systems include
$\eta '$ into two gluons
and the decays of the $\omega$ meson \cite{rob_1}.
Electroweak processes include couplings of the
axion.  Whether the changes in these couplings are
physically relevant can only be determined after detailed
calculation; what is certain is that they do change.
In a cosmological context, the mechanisms of electroweak baryogenesis
involve the effective coupling of a pseudoscalar field to the
Pontryagin density for the $SU(2)_L$ gauge field. In vacuum, this
coupling is directly related to the anomalous divergence of the baryon
current.  It was argued in ~\cite{dine} that this coupling must
be suppressed above the electroweak phase transition; presumably
this can be understood from our analysis.
Lastly, we note that
the 't Hooft anomaly matching conditions \cite{hooft}
strongly constrain the appearance of massless bound states
in confining theories.  Due to the failure
of the Sutherland-Veltman theorem, sec. IV,
these conditions are clearly much less restrictive at nonzero temperature.

If nothing else, perhaps this gives us a greater appreciation
of the wonder of the fermion axial anomaly.  While every other
anomalous decays changes in complicated
and detailed ways, that alone remains inviolate, always.

\bigskip
We thank W. Marciano 
for useful conversations.
This work is supported by a DOE grant at 
Brookhaven National Laboratory, DE-AC02-76CH00016.

\appendix
\section{WZW action}
\label{awzw}

In this appendix we review the Wess-Zumino-Witten (WZW) model
coupled to an external photon field.  
Along with establishing notation, this also enables us
to discuss a novel form of the WZW model, mentioned
recently~\cite{spence,creutztyt}, and to comment about 
the two currents in the two-flavor version of the WZW model.

For $n_f$ flavors, the model is constructed from a 
$n_f \times n_f$ unitary matrix $g$, $g^\dagger g = 1$.  In the
absence of gauge fields, the action is the sum
of two terms, ${\cal S} = {\cal S}_0 + {\cal S}_{wzw}$, with
${\cal S}_0 = \int d^4 \! x \, {\cal L}_0$ 
the usual action for a nonlinear sigma model, 
\begin{equation}
{\cal L}_{0} =  {f^2_\pi \over 4}\; tr\left 
( \partial_\alpha g^\dagger \partial_\alpha g 
\right) \; .
\label{eq:a.1}
\end{equation}
The generators of $SU(n_f)$ are the matrices $\lambda^a$,
normalized as $tr(\lambda^a \lambda^b) = 2 \delta^{a b}$.
This lagrangian is invariant under global
$SU(n_f)_\ell \times SU(n_f)_r$ unitary transformations,
$g(x) \rightarrow \Omega_\ell g(x) \Omega_r^\dagger$.
For example, under axial transformations, 
for which $\Omega_\ell = \Omega_r^\dagger$,
the corresponding conserved current is 
\begin{equation}
{\cal J}_{5,\alpha} = R_\alpha + L_\alpha = g^\dagger \partial_\alpha g 
+ (\partial_\alpha g) g^\dagger \; .
\label{eq:a.2}
\end{equation}
The second piece of the action is the Wess-Zumino-Witten term,
\begin{equation}
{\cal S}_{wzw} = 
- i {N_c\over 240 \pi^2} \int d^5 \!x \;
\varepsilon_{\alpha \beta \gamma \delta \sigma}\,
tr\left(R_\alpha R_\beta R_\gamma R_\delta R_\sigma 
\right ) \; , 
\label{eq:a.3}
\end{equation}
where the integral is over a five-dimensional region
whose boundary is four-dimensional spacetime.  

We wish to couple $g$ to a photon field $A_\alpha(x)$
in a gauge invariant manner.
For three flavors the charge matrix is
$Q = {\bf 1}/6 + \lambda_3/2$, 
and so we introduce the
covariant derivative,
$D_\alpha = \partial_\alpha + i e A_\alpha \left [Q,\cdot\right]$.
By construction, both $g$ and $D_\alpha g$
transform covariantly under local $U(1)$ gauge
transformations, $A_\alpha \rightarrow A_\alpha 
+\partial_\alpha \theta(x)$,
$\Theta(x) = exp(- i e Q \, \theta(x))$,
$g \rightarrow \Theta g \Theta^\dagger$,
$D_\alpha g \rightarrow \Theta (D_\alpha g) \Theta^\dagger$.

As usual, to make ${\cal L}_0$ gauge invariant 
we simply replace the ordinary
by the covariant derivative,
$\widetilde{{\cal L}}_0 =
f_\pi^2 tr|D_\alpha g|^2/4$.  We can also do this for
${\cal S}_{wzw}$ by replacing $R_\alpha$ with the gauge
covariant $\widetilde{R}_\alpha = g^\dagger D_\alpha g$.  
Using $\widetilde{R}_\alpha$ in ${\cal S}_{wzw}$ gives
something which is manifestly gauge invariant, but incomplete,
since it can be shown that the ensuing equations of 
motion depend on the fifth dimension.
In five dimensions, however, there are several other gauge
invariant terms which can be added, involving powers 
of the (abelian, gauge invariant) field strength, 
$F_{\alpha \beta} = \partial_\alpha A_\beta - \partial_\beta A_\alpha$.
The correct action then follows uniquely by
requiring that the equations of motion are independent of the
fifth dimension\cite{spence,creutztyt},
\begin{eqnarray}
\label{eq:a.4}
\widetilde{{\cal S}}_{wzw} 
&& \mbox{} = - i {N_c\over 240 \pi^2} \int d^5\! x \;
\varepsilon_{\alpha \beta \gamma \delta \sigma}\,
\left\{ tr\left( \widetilde{R}_\alpha 
\widetilde{R}_\beta \widetilde{R}_\gamma 
\widetilde{R}_\delta \widetilde{R}_\sigma \right) \right.
\nonumber\\ 
&& \mbox{} + 5 F_{\alpha \beta}\, tr\left( Q 
(\widetilde{L}_\gamma \widetilde{L}_\delta
\widetilde{L}_\sigma 
+  \widetilde{R}_\gamma 
\widetilde{R}_\delta \widetilde{R}_\sigma )\right)\nonumber\\
&& \mbox{} - 10 F_{\alpha \beta} F_{\gamma \delta} 
tr\left( Q^2(\widetilde{L}_\sigma + \widetilde{R}_\sigma)
\right.\nonumber\\
&&\mbox{} +
\left.\left.{1\over 2} Q g^\dagger Q D_\sigma g -
 {1\over 2} Q g Q D_\sigma g^\dagger \right)
 \right\} \; .
\end{eqnarray}

This expression is manifestly gauge invariant but not
obviously independent of the fifth dimension~\cite{footnote}.  
This is in contrast to the usual action, in which
the terms which couple to the gauge field are manifestly
four dimensional, but not evidently gauge invariant:
$\widetilde{{\cal S}}_{wzw} =
{\cal S}_{wzw} + \int d^4 \! x \, {\cal L}^A_{wzw}$, where 
\begin{equation}
{\cal L}^A_{wzw} =
- e N_c A_\alpha {\cal J}_\alpha
+ i e^2 N_c \varepsilon_{\alpha \beta \gamma \delta} \partial_\alpha
A_\beta A_\gamma {\cal K}_\delta \; ,
\label{eq:a.5}
\end{equation}
where
\begin{equation}
{\cal J}_\alpha
= \frac{1}{48 \pi^2} \; \varepsilon_{\alpha \beta \gamma \delta} \;
tr \left\{ Q(L_\beta L_\gamma L_\delta + R_\beta R_\gamma R_\delta) 
\right\} \; ,
\label{eq:a.6}
\end{equation}
and
\begin{eqnarray}
{\cal K}_\alpha &=& \frac{1}{24 \pi^2}
tr \left\{  Q^2(L_\alpha + R_\alpha) \right.
\nonumber\\
&& \left. + {1\over2} \left(
Q g Q g^\dagger L_\alpha + Q g^\dagger Q g R_\alpha \right) \right\} \; .
\label{eq:a.7}
\end{eqnarray}
The Lagrangian~\ref{eq:a.5} is the form given by Witten~\cite{witten}: it is
gauge invariant up to boundary terms. 
 
Henceforth we follow Brihaye, Pak, and Rossi~\cite{kayma} and 
restrict ourselves to two flavors.  Introducing the pion
field $\pi^a$,
\begin{equation}
\label{eq:a.8}
g = \exp\left( i\,{\pi^a \lambda^a\over f_\pi}\right) \; ,
\end{equation}
then for $SU(2)$
\begin{equation}
g  = cos\phi + i \, \frac{\pi^a \lambda^a}{ \sqrt{ \vec{\pi}^2 } } 
\; sin\phi \;\;\; ,\;\;\; 
\phi = \frac{\sqrt{\vec{\pi}^2}}{f_\pi} \; ,
\label{eq:a.9}
\end{equation}
$\vec{\pi}^2 = \pi^a \pi^a$.
The original lagrangian ${\cal L}_0$ becomes
\begin{equation}
{\cal L}_0 = 
\frac{1}{2} \frac{sin^2 \phi}{\phi^2} (\partial_\alpha \vec{\pi})^2
+ \frac{1}{2 f_\pi^2} \frac{(\phi^2 - sin^2\phi)}{\phi^4}
\left( \vec{\pi} \cdot \partial_\alpha \vec{\pi} \right)^2 \; .
\label{eq:a.10}
\end{equation}
The axial current of (\ref{eq:a.2}) is
\begin{eqnarray}
{\cal J}_{5,\alpha}^a &=& 
4 i \left( \frac{sin \phi \, cos \phi}{\phi} \; \partial_\alpha \pi^a
\nonumber \right. \\
&& \left. + \frac{(\phi - sin \phi \, cos \phi)}{f_\pi^2 \phi^3}
\; \vec{\pi} \cdot \partial_\alpha \vec{\pi} \; \pi^a \right) \; .
\label{eq:a.11}
\end{eqnarray}
In the WZW term, $S_{wzw}$ vanishes, while using the identity,
\begin{equation}
\varepsilon^{a b c} \varepsilon_{\alpha \beta \gamma \delta}
\left( \vec{\pi} \cdot \partial_\beta \vec{\pi} \pi^a 
- \frac{\vec{\pi}^2}{3} \partial_\beta \pi^a \right)
\partial_\gamma \pi^b \partial_\delta \pi^c = 0 \; ,
\label{eq:a.12}
\end{equation}
we find that
\begin{equation}
{\cal J}_\alpha = \frac{1}{72 \pi^2 f_\pi^3}
\frac{sin^2\phi}{\phi^2} \varepsilon^{a b c}
\varepsilon_{\alpha \beta \gamma \delta}
\partial_\beta \pi^a
\partial_\gamma \pi^b
\partial_\delta \pi^c \; .
\label{eq:a.13}
\end{equation}
The charge $Q= {\bf 1}/6 +\lambda_3$; in 
going from (\ref{eq:a.6}) to (\ref{eq:a.13}), it turns out that the 
contribution from the piece $\sim \lambda_3$ drops out.  This
demonstrates that ${\cal J}_\alpha$ is directly proportional to the
baryon current~\cite{witten}.  
Further, we find that
${\cal K}_\alpha$ is proportional to the axial current in
the isospin-$3$ direction, (\ref{eq:a.11}),
\begin{equation}
{\cal K}_\alpha = \frac{1}{96 \pi^2 f_\pi} {\cal J}^3_{5,\alpha} \; .
\label{eq:a.14}
\end{equation}
That for two flavors
${\cal J}_\alpha$ and ${\cal K}_\alpha$ are proportional
to the baryon and axial isospin currents
does not seem to have been recognized previously.  
Notice that ${\cal K}_\alpha$
is only proportional to the axial current for the original
lagrangian, ${\cal L}_0$; as is
discussed in sec~\ref{sec:adlerbardeen}, the complete axial
current includes contributions from the WZW term.
We do not know if ${\cal J}_\alpha$
and ${\cal K}_\alpha$ are equal to the analogous
currents for three or more flavors.

To conclude, we discuss how to compute the axial
current $J_{5, \mu}^3$ of 
sec.~\ref{sec:adlerbardeen} using the Noether construction.
The general form for the gauged axial current is
rather involved.  Since we compute perturbatively, however,
we just compute the current in the same way.
Under an infinitesimal axial rotation,
$\Omega_\ell = \Omega_r^\dagger = exp(i \omega)$,
the pion field transforms nonlinearly,
\begin{equation}
\pi^a \rightarrow \pi^a + f_\pi \omega^a
+ \frac{1}{3 f_\pi}(- \vec{\pi}^2 \, \omega^a + 
\vec{\pi}\cdot\vec{\omega} \, \pi^a ) + \ldots \; ,
\label{eq:a.15}
\end{equation}
to $\sim \omega$ and $\sim \pi^3$.
After performing such a transformation, the
requisite current is then the coefficient of
$\partial_\alpha \omega^3$.

In the present case, however, while the lagrangian is (of
course) gauge invariant, the lagrangian density need not be.
In particular, the lagragian density of (\ref{eq:a.5}) is
not gauge invariant; under a local gauge transformation,
it transforms by a total derivative.  To the
order at which we compute, we add the following term to
the lagrangian density:
$$
- \frac{e^2 N_c}{24 \pi^2} \;
\epsilon_{\alpha \beta \gamma \delta} 
\partial_\alpha \left( \pi^3 A_\beta \partial_\gamma A_\delta \right) 
$$
\begin{equation}
- \frac{i e N_c}{24 \pi^2} \;
\epsilon_{\alpha \beta \gamma \delta} 
\partial_\alpha \left(
A_\beta ( \pi^+ \partial_\gamma \pi^- -\pi^- \partial_\gamma \pi^+ ) 
\partial_\delta \pi^3 \right)  \; .
\; 
\label{eq:a.20}
\end{equation}
For example, the first piece contributes a Chern-Simons
term to the current.
With the addition of (\ref{eq:a.20}), the lagrangian density
is manifestly gauge invariant, and the axial current is
\begin{eqnarray}
\label{eq:a.19}
{J}_{5, \alpha}^3 &\approx&
f_\pi \partial_\alpha \pi^3
+ \frac{2}{3 f_\pi} \left(
\vec{\pi} \cdot \partial_\alpha \vec{\pi} \; \pi^3 
- \vec{\pi}^2 \; \partial_\alpha \pi^3\right)\nonumber\\
&+& \frac{i e N_c}{24 \pi^2 f_\pi^2} \;
\varepsilon_{\alpha \beta \gamma \delta} \;
\partial_\beta A_\gamma 
(\pi^+ \partial_\delta \pi^- - \pi^- \partial_\delta \pi^+)
\nonumber \\
&-& \frac{e^2 N_c}{12 \pi^2 }
\left( \frac{\pi^+ \pi^-}{f_\pi^2} \right)
\varepsilon_{\alpha \beta \gamma \delta} A_\beta \partial_\gamma A_\delta
+ \ldots 
\end{eqnarray}
When $A_\alpha =0$, 
$J_{5,\alpha}^3$ reduces to the axial 
vector current in the absence of
electromagnetism, ${\cal J}_{5,\alpha}^3$,
as can be verified by expanding (\ref{eq:a.11}).
The terms linear and quadratic in $A_\alpha$
agree with the calculation from the Wess-Zumino
consistency condition, (\ref{eq:5.1}).

This ambiguity in the construction of the axial
current is familiar.  
Instead of the gauge invariant, anomalous current
used in sec.~\ref{sec:ward}, by subtracting off a Chern-Simons
term, we can choose to work instead with a current
which is conserved but not gauge invariant.

\section{Hard thermal loops}
\label{ahtl}

We collect some results of hard thermal loops, with
minor differences in notation from previous work~\cite{htl}.
The simplest hard thermal loop is the integral 
of (\ref{eq:3.1}), ${\cal I}_T = T^2/12$.  
After that, there is $\delta \Pi^{\alpha \beta}(P)$ of
(\ref{eq:3.3}) and (\ref{eq:3.4}).
By definition the hard thermal
loop includes only only the terms $\sim T^2$
in the integral, in the limit of
soft external momentum $P \ll K \sim T$.
In this approximation, $\Pi^{\alpha\beta}(P)$ is tranverse:
\begin{equation}
\label{eq:b.1}
P^\alpha \,\delta\Pi^{\alpha \beta }(P) = 0 \; .
\end{equation}
One way of writing $\delta \Pi^{\alpha \beta }(P)$ is
\begin{equation}
\label{eq:b.2}
\delta \Pi^{\alpha\beta}(P) = 
2 n^\alpha n^\beta + 2 \int {d\Omega\over 4\pi} 
\; \omega \;
{\hat K^\alpha \hat K^\beta \over P\cdot \hat K}\;  , 
\end{equation}
where $n_\alpha = (1, \vec 0)$ and $\hat K = ( i, \hat{k})$;
$\hat{k}$ is a three vector of unit norm, $\hat{k}^2 = 1$, so
that $\hat{K}^\alpha$ is null, $\hat{K}^2 = 0$, and
$P\cdot \hat K = i p_0 + \vec p\cdot
\hat k = \omega + p \cos \theta$.
It is a dummy variable, in that 
one integrates over all directions of
$\hat{k}$, as $\int d \Omega_{\hat{k}}/(4 \pi)$.
In component form,
$$
\delta\Pi^{00}(P) = - 2 Q_1(z) \; ,
$$
$$
\delta\Pi^{0i}(P) = - 2 i \, z \, Q_1(z) \hat p^i \; , 
$$
$$
\delta\Pi^{ij}(P) =  2 z^2 Q_1(z) \hat{p}^i \hat{p}^j 
$$
\begin{equation}
- \frac{2}{5} \left(Q_3(z) - Q_1(z) - \frac{5}{3} \right)
 \, (\delta_{ij} - \hat p^i \hat p^j) \; , 
\label{eq:b.3}
\end{equation}
where $z = \omega /p$,
and the $Q_i(z)$ are Legendre functions of the second kind,
\begin{eqnarray}
\label{eq:b.4}
Q_1(z) &=& \frac{z}{2} \ln\left( {z +1 \over z -1 }\right) - 1 \; ,\\
Q_3(z) &=&  \frac{z (5 z^2 - 3)}{4} \ln\left ({z+ 1\over z-1}\right )
- \frac{5}{2} z^2 + \frac{2}{3} \; . \\
\end{eqnarray}

For $\pi^0 \rightarrow \gamma \gamma$, we need
the value of $\delta \Pi^{\alpha \beta}(P)$ 
near the light cone, $\omega \sim p^+$,
$$
\delta \Pi^{\alpha \beta} \sim
\delta^{\alpha \beta}
+ \frac{1}{n \cdot P} (n^\alpha P^\beta + P^\alpha n^\beta)
$$
\begin{equation}
- \left( \ln\left(\frac{2 p }{\omega - p}\right) - 1 \right)
\frac{P^\alpha P^\beta}{p^2} \; ,
\label{eq:b.5}
\end{equation}
where $n^\alpha = (1,\vec{0})$.

An equivalent but useful expression for (\ref{eq:b.2}) is 
\begin{eqnarray}
\label{eq:b.6}
\Pi^{\alpha\beta}(P) &=&  \int {d\Omega\over 4\pi} \left [ 
\delta^{\alpha\beta} + {\hat K^\alpha
\hat K^\beta \over (\hat K \cdot P)^2} P^2  \right. \nonumber\\
&&\mbox{} \left.- \; {P^\alpha \hat K^\beta + 
\hat K^\alpha P^\beta  \over (\hat K\cdot P)}\right ] \; ;
\end{eqnarray}
from which follows
\begin{equation}
\label{eq:b.7}
T \sum \int \frac{d^3 p}{(2 \pi)^3} 
 \; X_\alpha \; \delta \Pi^{\alpha \beta}(P) \; Y_\beta 
 = 
\end{equation}
$$
\int \!\! d^4 x \!\! \int \!\! \frac{d \Omega_{ \hat{k} } }{4 \pi} 
\, \!\! 
(\partial_\mu X_\alpha \!\! - \!\! \partial_\alpha X_\mu )
\frac{\hat{K}^\alpha\hat{K}^\beta}{- (\partial \cdot \hat{K})^2}
(\partial_\mu Y_\beta \!\! - \!\! \partial_\beta Y_\mu) 
$$
The abelian field strengths of the vector fields
$X^\alpha$ and $Y^\beta$ enter because $\delta \Pi^{\alpha \beta}$
is transverse, (\ref{eq:b.1}).

\section{$\pi \rightarrow \gamma\gamma$ in the real time formalism}
\label{sec:realtime}

We found in sec.~\ref{sec:lowt} that 
the contribution from fig.~(1.d) vanishes to order $T^2/f_{\pi}^2$.
In this appendix we check this result using the real time
approach.  In this instance, the real time method is simple
and not problematic.  It also shows that 
the diagram has no temperature dependent terms whatsoever:
the entire diagram is equal, identically,
to its value at zero temperature.

After dropping irrelevant terms proportional to $P_{1}^{\alpha}$,
fig.~(1.d) is proportional to the integral
\begin{eqnarray}
\epsilon_{\beta \gamma \delta \kappa} P_{2}^{\kappa} \int \frac{d^4K}{(2
\pi)^3} (2K - P_1)^{\alpha} K^{\gamma} (K-P_1)^{\delta} \nonumber \\
\left(\frac{n(\omega _k)}{(K-P_1)^2} \delta(K^2) +
\frac{n(\omega_{k-p_1})}{K^2} \delta((K-P_1)^2) \right) \nonumber \\
= - 2 \epsilon_{\beta \gamma \delta \kappa} P_{1}^{\delta} P_{2}^{\kappa}
\int \frac{d^4 K}{(2\pi)^3} \frac{K^{\alpha} K^{\gamma}}{K \cdot P_1}
\delta(K^2) n(\omega_k) \; ,
\end{eqnarray}
$\omega_k = k$, $\omega_{k-p_1} = |\vec{k}-\vec{p}_1|$.
The physical amplitude is obtained by contraction with
the polarization vectors for the two photons.  If $\vec{p}_1$
lies in the $z$-direction, say, then $\alpha$ must lie
along the $x$- or $y$-directions.  The angular $\phi$ integration
will then vanish unless $\gamma =\alpha$.  Thus the integral
reduces to
$$
\int \frac{d^4K}{(2 \pi)^3} \frac{\omega_k^2 \sin^2{\theta}}
{K \cdot P_1}  
\frac{n(\omega_k)}{\omega_k} (\delta(k_0 - \omega_K) +
\delta(k_0 + \omega_K)) \nonumber \\ 
$$
%
\begin{equation}
= \int \frac{d^3 k}{(2 \pi)^3} \frac{2 \cos{\theta}}{p_1^0}
n(\omega_k).
\end{equation}
This integral vanishes after integration over $\theta$;
note that this physical amplitude is free of any collinear
divergences.  This confirms our results obtained with the imaginary
time formalism.

\section{$\gamma\rightarrow \pi\pi\pi$ at low temperature}
\label{gamma3pi}

In this section we compute the one loop corrections
to the amplitude $\gamma \rightarrow \pi\pi\pi$
for soft, cool pions.
This provides another, less trivial, example of
hard thermal loops.  Corrections to the five dimensional 
Wess-Zumino-Witten term, (\ref{eq:a.3}),
have been computed in~\cite{pistyt2} using a background
field method; it would be interesting 
using this method to compute the one
loop corrections to the gauged WZW model, (\ref{eq:a.5}).

To one-loop order, the corrections to $\gamma \rightarrow \pi\pi\pi$
are those of figs.~(4.a) and~(4.b).

\begin{figure}[hbt]
\centerline{\epsfig{figure=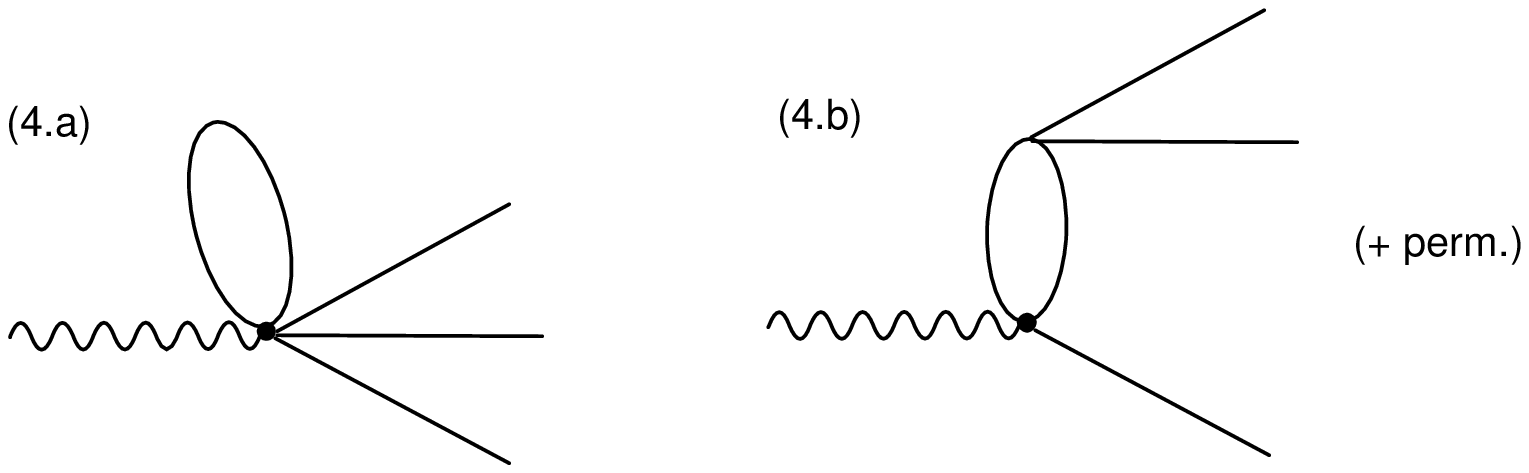,height=23mm}}
\end{figure}

Fig.~(4.a) is a tadpole diagram, which is most easily
computing by expanding (\ref{eq:a.13}) in the pion
field,
\begin{equation}
\label{eq:d.1}
{\cal L}_{\gamma\pi\pi\pi}
\simeq - {e N_c \over 72 \pi^2 f_b^3} 
\left(1 - \frac{\vec{\pi}^2}{3 f_\pi^2} \right)
\,A_\alpha\, \varepsilon^{a b c}
\varepsilon_{\alpha \beta \gamma \delta}
\partial_\beta \pi^a
\partial_\gamma \pi^b
\partial_\delta \pi^c \; ,
\end{equation}
and then compute as in (\ref{eq:2.4}).

At tree level, the amplitude for $\pi\pi\pi\gamma$ scattering is
\begin{eqnarray}
\label{eq:d.2}
{\cal M} =  \kappa_b\,
\varepsilon^{a b c}\, \varepsilon_{\alpha\beta\gamma\delta}\, 
\epsilon^\alpha P_\beta^a
P_\gamma^b P_\delta^c \; ,
\end{eqnarray}
where $P^a$ is the momentum of the pion with
isospin $a$, {\it etc}, and 
\begin{equation}
\label{eq:d.3}
 f_{b}^3\, \kappa_b = {i e N_c \over 72\pi^2} \; .
\end{equation}
From (\ref{eq:a.10}), the amplitude for the scattering
between four pions is
\begin{eqnarray}
\label{eq:d.4}
{\cal A}
 &=&  - {1\over f_b^2} \left[ \frac{}{}
\delta^{ab} \delta^{cd} P_{ab}^2  
+ \delta^{ac} \delta^{bd} P_{ac}^2 
+ \delta^{ad} \delta^{bc} P_{ad}^2 \right.\\ 
&-& \left. \frac{\sum P_i^2}{3} \left(
\delta^{ab} \delta^{cd} 
+ \delta^{ac} \delta^{bd} 
+ \delta^{ad} \delta^{bc} \right) \,\right] \; , \nonumber
\end{eqnarray}
where  $P_{ab}= P_a + P_b$, {\it etc}.

The contribution of fig.~(4.b) is
\begin{eqnarray}
\label{eq:d.5}
{\cal M}^{b} &=&  { 3 i\over f_b^2} \,{e N_c \over 72 \pi^2 f_b^3}\, 
 \varepsilon_{\alpha\beta\gamma\delta} 
\epsilon_\alpha P_\delta^c (P_\gamma^a + P^b_\gamma)
\times\\
&& 
\Gamma^{\beta\kappa}(P_{ab})(P^a_\kappa - P^b_\kappa)
+ \mbox{permutations} \; , \nonumber 
\end{eqnarray}
where $\Gamma^{\alpha\beta}(P_{ab})$ is the integral
of (\ref{eq:2.14}).

In the vacuum,
$\Gamma^{\alpha\beta}(P)  = \delta^{\alpha\beta} {\cal I}_0/2$,
up to terms $\sim P^\alpha P^\beta$ which drop out of the amplitude.
To ${\cal O}(P^4)$, 
\begin{equation}
\label{eq:d.7}
f_b^3\, \kappa_b = \left( 1 + (1 - 1 + 3)
{{\cal I}_0\over f^2}\right) {i e N_c \over 72\pi^2} \; .
\end{equation}
The $1$ comes from $Z_{\pi}^{3/2}$, fig.~(1.a) and (\ref{eq:2.5a}), 
the $-1$ from fig.~(4.a) and (\ref{eq:d.1}), the $+3$ from
fig.~(4.b) and~(\ref{eq:d.5}).
Using (\ref{eq:2.8}), then, to one loop order in vacuum~(\ref{eq:d.3})
renormalizes with no change in form,
\begin{equation}
\label{eq:d.8}
f_\pi^3 \,\kappa =  {i e N_c \over 72\pi^2} \; ,
\end{equation}
analogous to (\ref{eq:2.10}) and (\ref{eq:2.17}).

At low temperature, $\Gamma^{\alpha\beta}$ 
is replaced by $\delta \Gamma^{\alpha\beta}$ of 
(\ref{eq:3.3}) and (\ref{eq:3.4}).
Using~(\ref{eq:b.7}), we find that
the effective Lagrangian for $\gamma\rightarrow
\pi\pi\pi$ is similar to that for 
$\pi^0 \rightarrow \gamma \gamma$, (\ref{eq:3.9}).
One term is as at zero temperature, with $f_\pi$
replaced by $f_\pi(T)$, while the second is a hard thermal loop:
\begin{equation}
\label{eq:d.9}
{\cal L}_{\pi\pi\pi\gamma}(T) = -  
{e N_c \over 72\pi^2 f_\pi(T)^3}\,\,A_\alpha\,
\varepsilon^{a b c}
\varepsilon_{\alpha \beta \gamma \delta}
\partial_\beta \pi^a
\partial_\gamma \pi^b
\partial_\delta \pi^c
\end{equation}
$$
- \frac{T^2}{12 f_\pi^2}\, { e N_c\over 48 \pi^2 f_\pi^3}\,
\varepsilon^{abc} \int \frac{d \Omega_{ \hat{k} } }{4 \pi} \; 
 H^a_{\gamma \alpha}
\frac{ \hat{K}^\alpha \hat{K}^\beta}{- (\partial \cdot \hat{K})^2}
J^{b c}_{\gamma\beta} \; ,
$$
where 
\begin{eqnarray}
\label{eq:d.11}
H^a_{\alpha,\beta} &=& 
\partial_\alpha
(\varepsilon_{\beta\gamma\delta\kappa} 
F_{\gamma\delta} \partial_\kappa  \pi^a)
- (\alpha \leftrightarrow \beta) \; , \\
J^{b c}_{\alpha \beta} &=& 
\partial_\alpha \pi^b \partial_\beta \pi^c
- \partial_\beta \pi^b \partial_\alpha \pi^c \; . \nonumber
\end{eqnarray}
%


\begin{references}
\bibitem{estrada3}
$\Gamma^{\alpha \beta}(P)$, (2.13), equals 
$I^{\alpha \beta}(K)$, (5) of ref. \cite{estrada2}.  
The terms $\sim T^2$ in $\Gamma^{\alpha \beta}(P)$
define $\delta \Gamma^{\alpha \beta}(P)$, (3.4).
From fig.~(1.b) of ref. \cite{estrada2}, which is our fig.~(1.d),
in (12) they obtain 
%
$
I^{i j }(K) \sim + \delta^{i j} T^2/12 + \ldots
$
%
This does not agree with (\ref{eq:3.7}), for which the
coefficient of the term $\sim \delta^{i j}$ vanishes.
\bibitem{footnote}
This method also allows one to gauge an arbitary subgroup
$H$ of $SU(n_f) \times SU(n_f)$, whether or not $H$ is anomaly
free.  If the $H$ anomalies do not cancel, however, there is
a doubling of the effective number of degrees of freedom, 
similar to the doubling which occurs 
for fermions on a lattice~\cite{creutztyt}.
\end{references}
\end{document}